# GRB 171010A / SN 2017htp: a GRB-SN at $z$=0.33

A. Melandri[1]⋆, D. B. Malesani[2,3,33,34], L. Izzo[4,2], J. Japelj[5], S. D. Vergani[6,1],
P. Schady[7], A. Sagués Carracedo[8], A. de Ugarte Postigo[4], J. P. Anderson[22],
C. Barbarino[9], J. Bolmer[6], A. Breeveld[10], P. Calissendorff[32], S. Campana[1],
Z. Cano[4], R. Carini[11], S. Covino[1], P. D'Avanzo[1], V. D'Elia[12,11], M. della Valle[24,30],
M. De Pasquale[13], J. P. U. Fynbo[33], M. Gromadzki[23], F. Hammer[6],
D. H. Hartmann[17], K. E. Heintz[14], C. Inserra[29], P. Jakobsson[14], D. A. Kann[4],
J. Kotilainen[36,35], K. Maguire[25], N. Masetti[15,18], M. Nicholl[20,21], F. Olivares E.[27,28],
G. Pugliese[26], A. Rossi[15], R. Salvaterra[19], J. Sollerman[9], M. B. Stone[35],
G. Tagliaferri[1], L. Tomasella[31], C. C. Thöne[4], D. Xu[37], and D. R. Young[16]

*Authors affiliations are listed at the end of the paper*

**ABSTRACT**

The number of supernovae known to be connected with long-duration gamma-ray bursts is increasing and the link between these events is no longer exclusively found at low redshift ($z \lesssim 0.3$) but is well established also at larger distances. We present a new case of such a liaison at $z$ = 0.33 between GRB 171010A and SN 2017htp. It is the second closest GRB with an associated supernova of only three events detected by Fermi-LAT. The supernova is one of the few higher redshift cases where spectroscopic observations were possible and shows spectral similarities with the well-studied SN 1998bw, having produced a similar Ni mass ($M_{\rm Ni}$ = 0.33 ± 0.02 M$_\odot$) with slightly lower ejected mass ($M_{\rm ej}$ = 4.1 ± 0.7 M$_\odot$) and kinetic energy ($E_{\rm K}$ = 8.1 ± 2.5 × $10^{51}$ erg). The host-galaxy is bigger in size than typical GRB host galaxies, but the analysis of the region hosting the GRB revealed spectral properties typically observed in GRB hosts and showed that the progenitor of this event was located in a very bright HII region of its face-on host galaxy, at a projected distance of ∼ 10 kpc from its galactic centre. The star-formation rate (SFR$_{GRB}$ ∼ 0.2 M$_\odot$ yr$^{-1}$) and metallicity (12 + log(O/H) ∼ 8.15 ± 0.10) of the GRB star-forming region are consistent with those of the host galaxies of previously studied GRB-SN systems.

## 1 INTRODUCTION

The connection between "under-energetic" long-duration gamma-ray bursts (GRBs) and X-ray flashes (XRFs) with Type Ic core-collapse supernovae (SNe) is well established (Woosley & Bloom 2006; Cano et al. 2017). These highly stripped-envelope SNe are firmly associated with many low-redshift ($z \lesssim 0.3$) GRBs or XRFs both photometrically and spectroscopically (e.g. Galama et al. 1998; Hjorth et al. 2003; Malesani et al. 2004; Pian et al. 2006; Bufano et al. 2012; Melandri et al. 2012; Schulze et al. 2014; D'Elia et al. 2015; Izzo et al. 2019) since at those distances optical photometric and spectroscopic accurancy is adequate to identify several features in the SNe spectra.

At larger distances ($0.3 \lesssim z \lesssim 1$) the association with the supernova is often inferred from the detection of a re-brightening in the late afterglow optical light curve and supported by sporadic spectroscopic observations acquired near the putative supernova maximum brightness (e.g.

⋆ E-mail: andrea.melandri@inaf.it

Greiner et al. 2003; Zeh et al. 2004; Della Valle et al. 2006; Soderberg et al. 2006; Cano et al. 2011; Sparre et al. 2011; Jin et al. 2013; Xu et al. 2013; Melandri et al. 2014b).

GRB 171010A is one of the few border-line bright cases at $z \sim 0.3$ for which photometric and spectroscopic observations were possible. In this paper we present a comprehensive analysis of GRB 171010A, of the associated supernova SN 2017htp and its host galaxy. In Section 2 we present the available multi-band data; in Section 3 we discuss the results of our photometric and spectroscopic analysis and in Section 4 we present our conclusions.

Throughout the paper, distances are computed assuming a $\Lambda$ CDM-Universe with $H_0 = 71$ km s$^{-1}$ Mpc$^{-1}$, $\Omega_m = 0.27$, and $\Omega_\Lambda = 0.73$ (Larson et al. 2011; Komatsu et al. 2011). We use the convention $F_\nu \propto t^{-\alpha}\nu^{-\beta}$. Magnitudes are in the AB system, not corrected for Galactic extinction $E_{B-V} = 0.13$ mag (Schlafly & Finkbeiner 2011), and errors are reported at a $1\sigma$ confidence level.

## 2 OBSERVATIONS

At 19:00:50.6 UT (=$T_0$ hereafter) on 2017, October 10, GRB 171010A was detected with the *Fermi*-LAT and *Fermi*-GBM instruments (Omodei et al. 2017; Poolakkil & Meegan 2017). A few hours later a counterpart was identified near a faint galaxy both in the NIR/optical (Thorstensen & Halpern 2017; Izzo & Malesani 2017) and in the X-rays (D'Aí et al. 2017) at coordinates RA = 04:26:19.46, Dec = −10:27:45.9 (J2000).

The burst light curve showed a multi-peaked structure in $\gamma$-rays with total duration $T_{90} \sim 160$ s[1] (Frederiks et al. 2017). The isotropic energy release ($E_{iso} \sim 2 \times 10^{53}$ erg at $z$=0.33) is consistent with the bright half of the isotropic energy distribution for long GRBs (Nava et al. 2012). Together with the rest-frame peak energy of the time-integrated spectrum ($E_{p,i} \sim 230$ keV), it makes GRB 171010A consistent with the $E_{p,i}$-$E_{iso}$ correlation of long GRBs (Amati 2006).

### 2.1 X-ray Imaging

The bright X-ray afterglow of GRB 171010A was detected by the Neil Gehrels *Swift* Observatory (*Swift* hereafter) about 6.5 hr after the burst event (Evans 2017; D'Aí et al. 2017). The source was detected in the X-ray band with an initial 0.3-10 keV absorbed flux of $\sim 3.7 \times 10^{-11}$ erg cm$^{-2}$ s$^{-1}$ and still above the detection threshold up to $\sim$20 days after the burst event. The light curve can be described with a single power-law decay $\alpha_X = 1.45 \pm 0.04$, with the typical small variability seen in several GRBs (e.g. Melandri et al. 2014a).

### 2.2 Optical Imaging

We observed the field of GRB 171010A with the 0.6-m iTelescope (iT), the Copernico 1.82-m telescope operated by INAF Osservatorio Astronomico di Padova (Asiago,

---

[1] Konus-WIND data show that the main pulse was preceded by a weaker emission starting from $T_0$−150 s and a further episode of emission started at about $T_0$+250 s

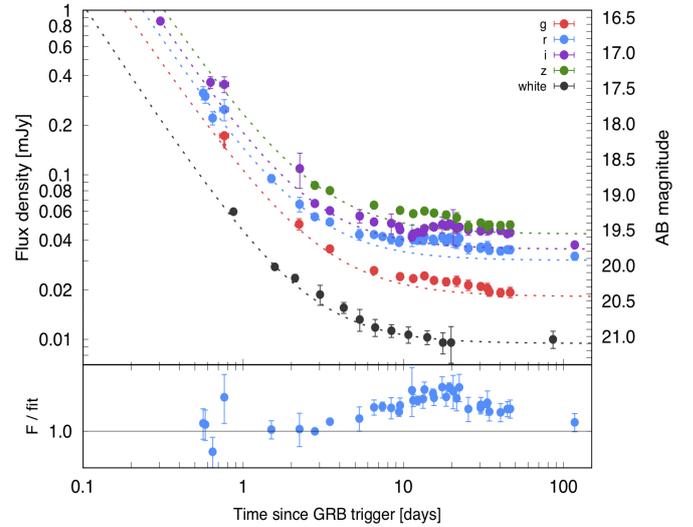

**Figure 1.** Observed optical light curve of GRB 171010A. For the sake of clarity we show only the $r$ filter in the residuals plot (bottom panel). Dashed lines represent the sum of the afterglow and host galaxy contribution. The deviation at late time (bump at $\sim$10 days) from the general decay is clearly due to the contribution of the associated supernova, SN 2017htp.

Mt. Ekar), the 2.2-m telescope at the Centro Astronómico Hispano-Alemán (CAHA), the 2.2-m MPG/ESO telescope (GROND), the 2.56-m Nordic Optical Telescope (NOT), the 3.6-m New Technology Telescope (NTT), the 3.6-m Telescopio Nazionale Galileo (TNG), the 8.2-m Very Large Telescope (VLT), and the 8.4-m Large Binocular telescope (LBT) for several imaging epochs (Table 1 and Fig. 1).

When *Swift* initially slewed to observe the afterglow of GRB 171010A, the position was not well known and the source was outside the field of view of the *Swift* Ultraviolet Optical Telescope (UVOT). Subsequent observations were made with a better pointing; in total 23 segments of data were taken. UVOT images were screened manually for source trailing or bad source positions (close to an edge, corner or outside the field of view) and any affected images were discarded, resulting in 203 exposures in the White filter ($\lambda_{min}^{Wh} \sim 1600$ Å, $\lambda_{min}^{Wh} \sim 7500$ Å), 5 exposures in $U$ filter and 10 exposures in $V$ filter. The White data cover the time period between 69 and 7278 ks after the trigger (Table 2 and Fig. 1).

### 2.3 Optical Spectroscopy

We observed the optical afterglow of GRB 171010A with the EFOSC2 instrument mounted onto the ESO/NTT 3.6-m telescope located in La Silla (Chile) and with the X-shooter spectrograph mounted onto the VLT/UT2 8.2-m telescope located in Paranal (Chile). The diary of the observations is shown in Table 3. The wavelength range (resolution) of our spectra is $\sim 3400 - 9300$ Å (R$\sim$1000) and $\sim 3000 - 10000$ Å (R$\sim$6000) for the NTT and the X-shooter observation, respectively.

**Table 1.** Observations log for GRB 171010A - SN 2017htp. Magnitudes are in the AB system, not corrected for Galactic extinction. $t-T_0$ refers to mid-point of the exposure.

| $t-T_0$ (days) | $t_{exp}$ (s) | Filter | Mag (err) | Telescope | $t-T_0$ (days) | $t_{exp}$ (s) | Filter | Mag (err) | Telescope |
|---|---|---|---|---|---|---|---|---|---|
| 2.236 | 720 | g | 20.08 (0.06) | Asiago | 117.38 | 1800 | r | 20.48 (0.06) | LBT |
| 3.478 | 1592 | g | 20.52 (0.03) | GROND | 0.626 | 900 | i | 17.75 (0.09) | iT |
| 6.556 | 1593 | g | 20.85 (0.05) | GROND | 2.256 | 720 | i | 19.06 (0.26) | Asiago |
| 9.535 | 2390 | g | 20.94 (0.05) | GROND | 3.478 | 1592 | i | 19.70 (0.03) | GROND |
| 11.528 | 2374 | g | 20.97 (0.04) | GROND | 5.332 | 1800 | i | 19.78 (0.10) | CAHA |
| 13.539 | 4153 | g | 20.93 (0.05) | GROND | 6.556 | 1593 | i | 19.87 (0.05) | GROND |
| 15.544 | 2180 | g | 21.00 (0.05) | GROND | 8.400 | 360 | i | 19.80 (0.12) | NTT |
| 18.495 | 2418 | g | 21.02 (0.06) | GROND | 9.401 | 600 | i | 19.87 (0.03) | NOT |
| 21.507 | 1216 | g | 21.01 (0.08) | GROND | 9.535 | 2390 | i | 19.99 (0.05) | GROND |
| 25.451 | 2403 | g | 21.07 (0.08) | GROND | 11.328 | 900 | i | 20.10 (0.04) | NOT |
| 30.457 | 780 | g | 21.09 (0.05) | GROND | 11.528 | 2374 | i | 20.05 (0.04) | GROND |
| 33.475 | 776 | g | 21.12 (0.08) | GROND | 12.311 | 900 | i | 20.03 (0.03) | NOT |
| 34.434 | 2410 | g | 21.17 (0.06) | GROND | 13.294 | 900 | i | 20.03 (0.04) | NOT |
| 40.409 | 2415 | g | 21.18 (0.06) | GROND | 13.539 | 4153 | i | 19.97 (0.05) | GROND |
| 46.411 | 2396 | g | 21.18 (0.08) | GROND | 15.321 | 1200 | i | 19.94 (0.02) | NOT |
| 1.500 | 720 | r | 19.30 (0.06) | NTT | 15.544 | 2180 | i | 19.95 (0.05) | GROND |
| 2.247 | 720 | r | 19.69 (0.11) | Asiago | 17.486 | 900 | i | 19.91 (0.12) | NTT |
| 3.477 | 1592 | r | 19.96 (0.02) | GROND | 18.495 | 2418 | i | 19.93 (0.06) | GROND |
| 5.310 | 900 | r | 20.15 (0.08) | CAHA | 19.383 | 1500 | i | 19.90 (0.02) | NOT |
| 6.556 | 1593 | r | 20.15 (0.05) | GROND | 20.461 | 600 | I | 19.92 (0.26) | VLT |
| 7.406 | 600 | r | 20.18 (0.03) | NOT | 21.507 | 1216 | i | 19.97 (0.08) | GROND |
| 8.394 | 360 | r | 20.22 (0.07) | NTT | 22.356 | 1500 | i | 19.95 (0.03) | NOT |
| 9.392 | 600 | r | 20.27 (0.03) | NOT | 25.451 | 2403 | i | 19.99 (0.08) | GROND |
| 9.535 | 2390 | r | 20.23 (0.05) | GROND | 30.372 | 900 | i | 20.01 (0.03) | NOT |
| 11.315 | 900 | r | 20.26 (0.06) | NOT | 30.457 | 780 | i | 20.00 (0.05) | GROND |
| 11.528 | 2374 | r | 20.23 (0.04) | GROND | 33.284 | 900 | i | 19.97 (0.02) | TNG |
| 12.298 | 900 | r | 20.24 (0.03) | NOT | 33.475 | 776 | i | 19.99 (0.08) | GROND |
| 13.282 | 900 | r | 20.24 (0.06) | NOT | 34.434 | 2410 | i | 19.99 (0.06) | GROND |
| 13.539 | 4153 | r | 20.18 (0.05) | GROND | 40.409 | 2415 | i | 20.00 (0.06) | GROND |
| 15.302 | 1500 | r | 20.22 (0.03) | NOT | 44.597 | 600 | i | 20.05 (0.05) | LBT |
| 15.544 | 2180 | r | 20.25 (0.05) | GROND | 46.411 | 2396 | i | 20.03 (0.06) | GROND |
| 17.476 | 900 | r | 20.19 (0.07) | NTT | 117.38 | 1800 | i | 20.22 (0.05) | LBT |
| 18.495 | 2418 | r | 20.26 (0.06) | GROND | 3.478 | 1592 | z | 19.33 (0.03) | GROND |
| 19.364 | 1500 | r | 20.20 (0.03) | NOT | 6.556 | 1593 | z | 19.55 (0.05) | GROND |
| 20.455 | 600 | R | 20.23 (0.11) | VLT | 9.535 | 2390 | z | 19.63 (0.05) | GROND |
| 21.507 | 1216 | r | 20.28 (0.08) | GROND | 11.528 | 2374 | z | 19.68 (0.04) | GROND |
| 22.341 | 900 | r | 20.21 (0.09) | NOT | 13.539 | 4153 | z | 19.64 (0.05) | GROND |
| 25.451 | 2403 | r | 20.36 (0.08) | GROND | 15.544 | 2180 | z | 19.67 (0.05) | GROND |
| 30.353 | 1500 | r | 20.36 (0.04) | NOT | 18.495 | 2418 | z | 19.70 (0.06) | GROND |
| 30.457 | 780 | r | 20.34 (0.05) | GROND | 21.507 | 1216 | z | 19.74 (0.08) | GROND |
| 33.272 | 900 | r | 20.34 (0.02) | TNG | 25.451 | 2403 | z | 19.87 (0.08) | GROND |
| 33.475 | 776 | r | 20.33 (0.08) | GROND | 30.457 | 780 | z | 19.82 (0.05) | GROND |
| 34.434 | 2410 | r | 20.39 (0.06) | GROND | 33.475 | 776 | z | 19.87 (0.08) | GROND |
| 40.409 | 2415 | r | 20.40 (0.06) | GROND | 34.434 | 2410 | z | 19.85 (0.06) | GROND |
| 44.597 | 600 | r | 20.38 (0.05) | LBT | 40.409 | 2415 | z | 19.86 (0.06) | GROND |
| 46.411 | 2396 | r | 20.38 (0.06) | GROND | 46.411 | 2396 | z | 19.85 (0.06) | GROND |

## 3 RESULTS

### 3.1 Photometric analysis

The images of all instruments were reduced following standard methods and we performed aperture photometry using the SkyCat-GAIA package[2]. We calibrated the optical *griz* data using a common set of selected catalogued stars detected in the field of view[3].

For UVOT white band data, which has a point spread function of ∼ 2 − 3 arcsec, a circular aperture of 3 arcsec radius, centered on the position given by the radio detection (Laskar et al. 2017) was used to measure the source count rate. Standard practice for UVOT photometry is to use a 5 arcsec radius aperture (Poole et al. 2008) but the smaller aperture was chosen to reduce contamination by the host galaxy, and an aperture correction was subsequently applied. The background was measured using a nearby source-free circular aperture of radius 53 arcsec. The measurements were made using the standard *Swift* FTOOL task UVOT-PRODUCT which sums up consecutive exposures to obtain a S/N ratio above a chosen limit (in this case 3), and corrects for coincidence loss, the time-dependent sensitivity loss and large scale sensitivity variations over the detector (Breeveld et al. 2010).

---
[2] http://star-www.dur.ac.uk/~pdraper/gaia/gaia.html
[3] https://catalogs.mast.stsci.edu/panstarrs/

**Table 2.** *Swift*/UVOT observation log for GRB 171010A-SN 2017htp. Magnitudes are in the white filter, in the AB system and they are not corrected for Galactic extinction.

| $t_{start}$ (s) | $t_{stop}$ (s) | $t_{exp}$ (s) | Mag (err) |
|---|---|---|---|
| 69337.7 | 80975.9 | 4818.4 | 19.81 (0.04) |
| 132777.8 | 139190.5 | 1929.9 | 20.65 (0.06) |
| 173477.4 | 190730.5 | 1505.1 | 20.72 (0.06) |
| 260605.6 | 261050.6 | 431.9 | 20.97 (0.13) |
| 346270.8 | 386990.6 | 3077.0 | 21.17 (0.06) |
| 460634.8 | 461930.5 | 1267.8 | 21.45 (0.10) |
| 557683.3 | 593750.5 | 4986.7 | 21.57 (0.07) |
| 683787.6 | 771590.6 | 11505.9 | 21.62 (0.05) |
| 850168.3 | 1006370.5 | 9110.8 | 21.68 (0.06) |
| 1115335.3 | 1321790.6 | 15454.0 | 21.72 (0.05) |
| 1402929.3 | 1648420.0 | 6185.0 | 21.80 (0.07) |
| 1711448.6 | 1723421.1 | 1511.1 | 21.80 (0.12) |
| 7277537.4 | 7572290.5 | 10332.0 | 21.75 (0.06) |

**Table 3.** Log of the spectroscopic observations. The second column refers to days from the *Fermi*-LAT trigger.

| Date | Time (UT) | $t - T_0$ (days) | Instrument | Exp. time (s) |
|---|---|---|---|---|
| 2017-10-12 | 07:05:01 | 1.503 | NTT/EFOSC2 | 2×1800 |
| 2017-10-19 | 04:58:01 | 8.415 | NTT/EFOSC2 | 2×2700 |
| 2017-11-01 | 05:18:39 | 21.430 | VLT/X-shooter | 4×1200 |

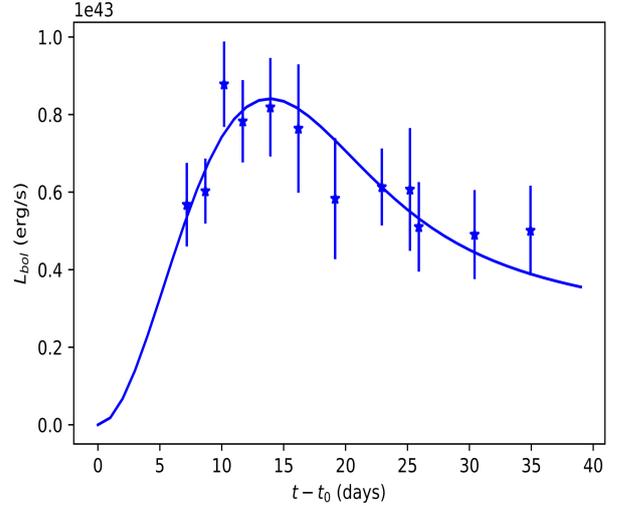

**Figure 2.** Quasi-bolometric (*griz*) rest-frame light curve of SN 2017htp fitted using the radioactive heating model (Arnett 1982) assuming fully trapped γ-rays. The model fits well to the data with a reduced $\chi^2$ of ~0.42 for 10 degrees of freedom. The model requires a nickel mass of $M_{Ni} = 0.33 \pm 0.02\ M_\odot$, ejecta mass of $M_{ej} = 4.1 \pm 0.7\ M_\odot$ and a kinetic energy of $E_K = 8.1 \pm 2.5 \times 10^{51}$ erg. For the analysis we have used the peak photospheric velocity derived from the spectral analysis of $v_{ph} = 14000 \pm 1000$ km s$^{-1}$ and a grey optical opacity of $k = 0.07$ cm$^2$ g$^{-1}$.

Next, the calibrated magnitudes reported in Table 1 were corrected for the Galactic absorption along the line of sight ($E_{B-V} = 0.13$ mag; Schlafly & Finkbeiner 2011) and converted into flux densities (Fukugita et al. 1996).

In Fig. 1 we show the temporal evolution of the optical afterglow of GRB 171010A. In this figure we include also data from public GCNs (Thorstensen & Halpern 2017; Guidorzi 2017; Harita 2017; Watson et al. 2017). We fitted the optical light curves with a single power-law decay plus a constant, in order to take into account the flattening of the light curves at late times, clearly due to the host-galaxy contribution. The early time decay index (for t<3 days) is similar for all the optical bands ($\alpha = 1.42 \pm 0.05$), and it is consistent with the temporal decay observed in the X-ray band (see Sec. 2.1). Moreover, a clear deviation of the general decay is visible in the *griz* filters after ~ 10 days, revealing the presence of the associated SN 2017htp, which is confirmed by the spectral analysis (Sec. 3.2).

In order to highlight the supernova contribution we subtracted the afterglow and host-galaxy contribution from the data, obtaining the SN 2017htp light curve for each filter. We then built the quasi-bolometric light curve of SN 2017htp (Fig. 2) and compared it with other well-studied broad-lined supernovae associated with low redshift GRBs (Fig. 3).

Typical *Fermi*-LAT afterglows are bright, energetic and distant events (Ajello et al. 2019), and only a few cases (among nearly 200 detected GRBs) were close enough to allow for an observational campaign in search of the associated supernova emission. Notably, GRB 171010A is the second closest GRB with an associated supernova of a total of three events detected by *Fermi*-LAT within the LAT catalog (Ajello et al. 2019). Previous events were GRB 130702A-SN 2013dx at $z = 0.145$ (D'Elia et al. 2015; Toy et al. 2016; Volnova et al. 2017) and GRB 130427A-SN 2013cq at $z = 0.34$ (Xu et al. 2013; Maselli et al. 2014; Melandri et al. 2014b; Becerra et al. 2017).

The physical mechanism powering the lightcurves of all SN types, therefore including GRB-SNe is the radioactive decay of Nickel and Cobalt into Iron. We have fitted the *griz* bolometric light curve of SN 2017htp with the radioactive heating model (Arnett 1982). In this model, the γ-rays produced in the decay are fully trapped by the SN ejecta, thermalizing it and producing the supernova emission. The free parameters are the effective diffusion timescale and the initial mass of nickel. The peak photospheric velocity derived from the spectral analysis is $v_{ph} = 14000 \pm 1000$ km s$^{-1}$ (see Sec. 3.2). We assumed an optical opacity of $k = 0.07$ cm$^2$ g$^{-1}$ and we obtain a nickel mass of $M_{Ni} = 0.33 \pm 0.02\ M_\odot$, an ejecta mass of $M_{ej} = 4.1 \pm 0.7\ M_\odot$ and a kinetic energy of $E_K = 8.1 \pm 2.5 \times 10^{51}$ erg. These values for the nickel and the ejected masses are in agreement with what was found for the other two *Fermi*-LAT GRB-SNe, while the estimated kinetic energy is at least a factor 5 and 4 fainter than what was observed for SN 2013cq and SN 2013dx, respectively (Melandri et al. 2014b; D'Elia et al. 2015). More generally the properties found for SN 2017htp are consistent with the average values found for the GRB-SNe population (Cano et al. 2017).

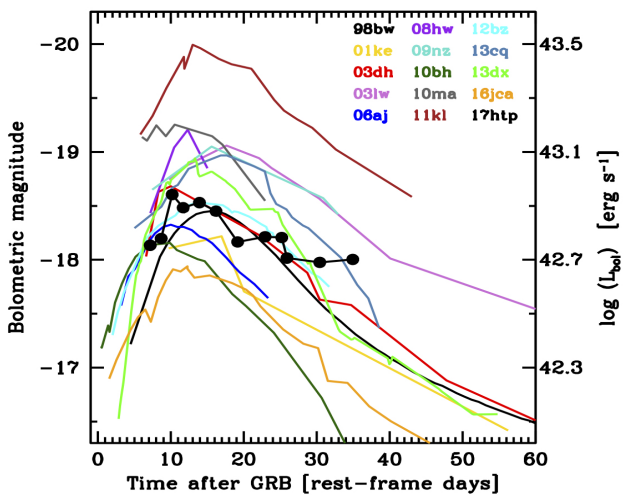

**Figure 3.** Comparison of the optical bolometric luminosity of SN 2017htp (black points) with a sample of well-studied GRB-SNe.

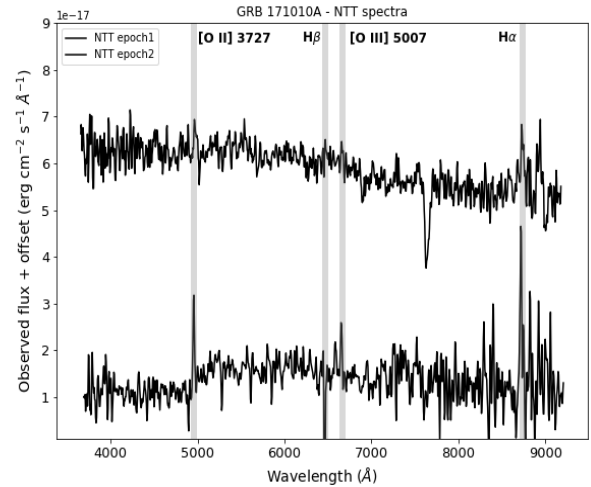

**Figure 4.** The NTT spectra detection of emission lines of H$\alpha$, H$\beta$, [O III] $\lambda$5007 and [O II] $\lambda\lambda$3727/3729 at the common redshift of $z = 0.33$.

### 3.2 Spectral analysis

The first spectrum was obtained 1.5 days after the GRB detection with the NTT/EFOSC2 instrument with two exposures of 1800 seconds each. The total spectrum shows a continuum described by a power-law function with spectral index $\beta = 1.33^{+0.10}_{-0.28}$ over which we detected emission lines of H$\alpha$, H$\beta$, [O III] $\lambda$5007 and [O II] $\lambda\lambda$3727/29 at the common redshift of $z = 0.33$ (Fig. 4). These emission lines are detected also in the second NTT spectrum acquired 8.4 days after the burst. No clear absorption features are detected. The lack of detection of absorption features in high S/N spectra is strongly suggestive of a low redshift for the afterglow.

We obtained a spectrum with the VLT, using the X-shooter spectrograph, about 21.5 days after the GRB. Spectroscopic observation covered both the host nucleus and the afterglow position (Fig. 7), revealing the presence of faint continuum emission at the GRB location (de Ugarte Postigo et al. 2017). This could be due to an associated supernova, as expected for long GRBs at this redshift, or to an underlying star forming region, or a combination of both. The spectral shape is curved and redder at the nucleus location, but the extinction is low as inferred from the Balmer decrement. This opens the possibility of excess light from the supernova. In the attempt to isolate its contribution, we then estimated the host contamination by rescaling the nucleus spectrum assuming it contributes the entire flux at 3500 Å (rest frame), where GRB-associated SNe have negligible emission (Hjorth et al. 2003; Mazzali et al. 2003), due to metal line blanketing.

By comparing the residual contribution with known SN-GRB templates, we find a good match with other GRB-associated SNe, such as SN 1998bw at 6 days post maximum (Patat et al. 2001), and SN 2006aj at 5 days post maximum (Pian et al. 2006) (see Fig. 5). In fact, at this epoch after the GRB trigger, we expected the SN to be around or just few days after the peak phase. A more detailed analysis shows the presence of Mg II, Fe II and Si II absorptions (Fig. 6) at expansion velocities of $v_{exp} \sim 13,000 - 15,000$ km s$^{-1}$, which are transitions typically observed in broad-lined Type-Ic SNe.

### 3.3 Host galaxy properties

We have imaged the host galaxy of GRB 171010A with the LBT in the $r$ and $i$ SDSS filters. The host is an almost face-on galaxy, with the main axis oriented along the direction of the X-shooter slit (see Fig. 7). We measure a major axis length of $r_{maj} = 5.3''$ and an axis ratio of 0.9. For a redshift of $z = 0.33$ this corresponds to a diameter of $\sim 25$ kpc. Following Japelj et al. 2018 we used both GALFIT (Peng et al. 2010) phenomenological fitting of the host's brightness as well as SExtractor (v2.19.5; Bertin & Arnouts 1996) to measure the galaxy size, obtaining consistent values of $r_{50}$ $(r_{90}) = 1.33''$ $(2.63'')$, corresponding to 6.6 (12.9) kpc. Assuming the GRB position as measured by the VLA afterglow observation (Laskar et al. 2017), we measure the offset with respect to the barycenter of the galaxy to be $1.4''$, which corresponds to a distance of $\sim 6.8$ kpc. Japelj et al. 2018 report and compare the properties of the host galaxies of Type Ic-BL SNe with and without an associated GRB. Their samples have $z < 0.2$. Despite being at a larger redshift, the host galaxy of GRB 171010A is bigger than all the hosts considered in Japelj et al. 2018 and its absolute magnitude at 4600 Å (corresponding to the observed $r$ band) $M = -20.8$ is at the very bright end of the distribution of Ic-BL host galaxies. On the other hand, the normalized offset of the SN position within the host corresponds to 1.0, which is typical of offsets of Type Ic-BL SNe with associated GRB.

In our X-shooter observations, we have positioned the slit such that it covers the GRB position and the entire host galaxy along the major axis. This allows us to study the properties of both the GRB 171010A region and its host as well (Fig. 7). The X-shooter spectrum of both sources is characterised by the presence of bright Balmer and nebular lines typically observed in GRB hosts, such as the [O III] $\lambda\lambda 4960,5007$, [O II] $\lambda\lambda 3727/3729$, [N II] $\lambda\lambda 6548/6584$ and [S II] $\lambda\lambda 6717/6732$. In the 2D spectrum, these lines show the typical shape due to the projection on the line of sight of the galaxy rotation velocity, see Fig. 8. We also note in the 2D spectrum that there is a bright H$\alpha$ emission at the star-forming region including the GRB position, suggesting that the GRB progenitor star was located in a very bright H II region of the host.

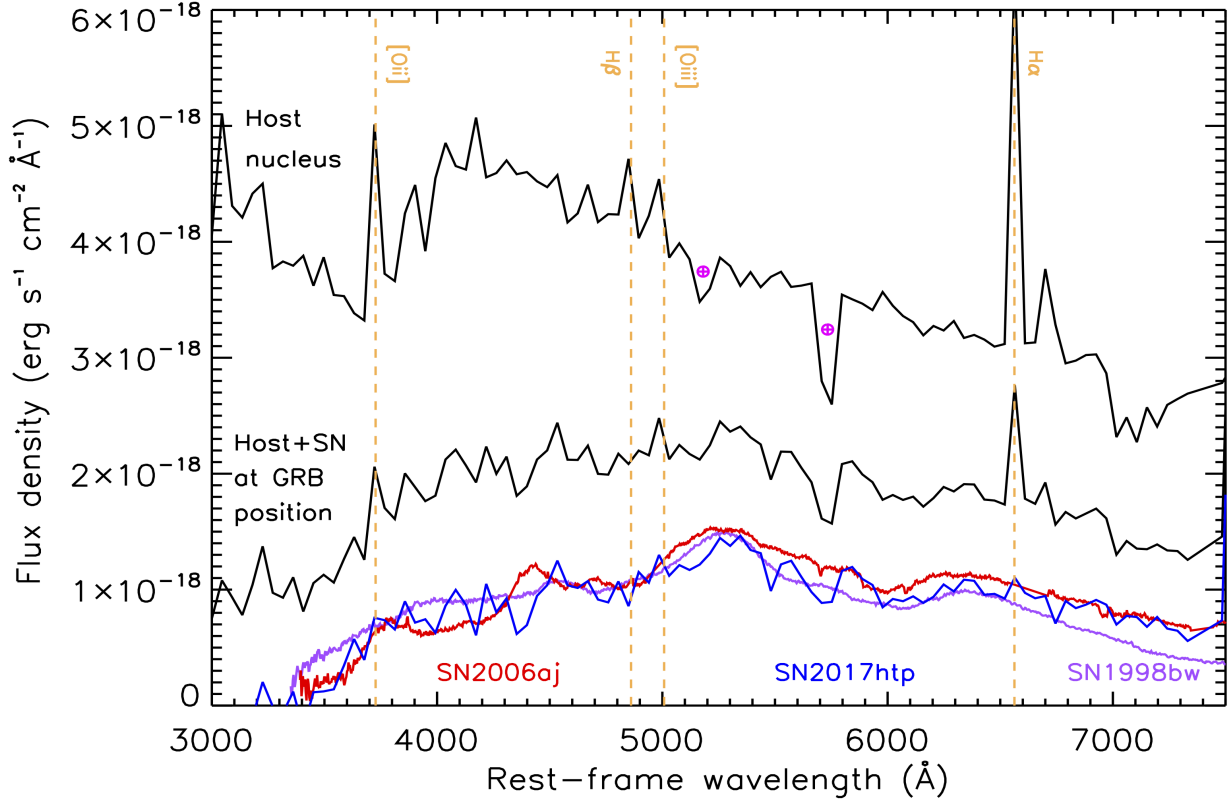

**Figure 5.** Residual contribution of SN 2017htp (blue) compared with known SN/GRB templates (SN 1998bw in purple and SN 2006aj in red, at 6 and 5 days post maximum, respectively). The flux of the host nucleus (top spectrum) has been divided by a factor 1.2 for clarity. The ⊕ symbols mark the telluric A and B absorption bands, while vertical dashed lines mark some emission lines detection.

We extracted two spectra, the first corresponding to the bright knot including the GRB region, using an extraction aperture of the spectrum of 1.3″, while the second corresponds to the emission of the host galaxy. This operation was possible given the excellent seeing (0.65″) at the epoch of the observations. After correcting for the Galactic reddening (assuming $E_{B-V} = 0.13$ mag, obtained from Schlafly & Finkbeiner 2011) we noted that the Balmer decrement is almost consistent with zero. Therefore we did not correct for any intrinsic host absorption. This suggests that the GRB progenitor was located almost face on with respect to our line of sight. From the spectra we have measured emission line fluxes in order to estimate some of the physical properties (star-formation rate, metallicity and the ionisation level) of the GRB environment as well as of its host galaxy, see Table 4.

The star-formation rate (SFR) was estimated using the H$\alpha$ emission line indicator (Kennicutt et al. 1994), which is a good one for the SFR in presence of typical nebular physical conditions ($T_e \approx 10,000$ K and case B recombination; Osterbrock 1989), and assuming a Chabrier initial mass function (Chabrier 2003). We obtain for the GRB region $SFR_{GRB,H_\alpha} = 0.19 \pm 0.04\ M_\odot$ yr$^{-1}$ while for the remaining part of the host galaxy spectrum we find $SFR_{host,H_\alpha} = 0.88 \pm 0.20\ M_\odot$ yr$^{-1}$. A similar analysis can be done using the [O II] $\lambda\lambda$ 3727/3729 doublet (Kewley et al. 2004): we obtain results that are consistent within the errors with the H$\alpha$ indicator, $SFR_{GRB,[OII]} = 0.24 \pm 0.07\ M_\odot$ yr$^{-1}$.

We furthermore computed a specific SFR, defined as SFR per unit luminosity (SFR / (L/L*); see e.g. Christensen et al. 2004). Luminosity-weighted SFR allows us to evaluate the relative strength of star-formation at the GRB site. We used the LBT *r*-band image of the host galaxy (see Section 2.2 and Fig. 7) to measure the luminosity in the region covered by the slit of the X-shooter spectrograph during the observation. The luminosity was measured both for the GRB and host region. We make sure to measure each luminosity in the same region from which we extracted the spectra (see Fig. 7). As a reference luminosity $L^*$ we take the value of $M_B = -20.4$, corresponding to the break value in the Schechter luminosity function of the $z = 0.3$ blue galaxy population (Faber et al. 2007). Using the measured H$\alpha$-based SFRs we estimate sSFR$_{GRB}$ ~ 3.9 and

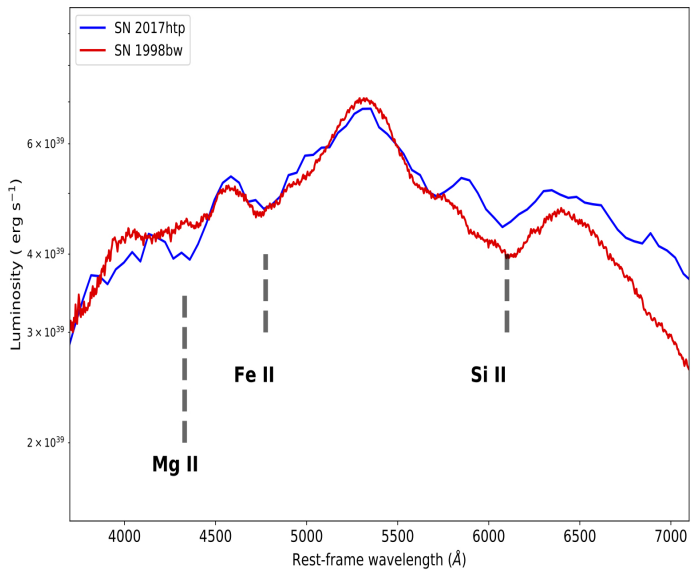

**Figure 6.** The spectrum of SN 2017htp, re-scaled in luminosity by a factor of 6 and smoothed with a boxcar kernel of width $w = 5$ Å, compared with the spectrum of SN 1998bw at similar phase. The very good match, and the identifications of Si II $\lambda6355$, Mg II $\lambda4481$ and the Fe II multiplet 42 at similar expanding velocities of $v_{exp} \sim 13,000 - 15,000$ km s$^{-1}$, confirms the presence of an underlying SN.

sSFR$_{host} \sim 1.6$ $M_\odot$ yr$^{-1}$/(L/L$^*$). We checked that these values are not strongly influenced by varying seeing conditions. Our results show that the GRB occurred in a region of enhanced star-formation compared to the rest of the host galaxy.

Similarly, we used emission lines to estimate the metallicity. In particular we used the N2 and O3N2 indicators that use the Balmer H$\alpha$ and H$\beta$ lines and the [N II] $\lambda6584$ and [O III] $\lambda5007$ nebular lines to provide an indication of the $12 + \log(O/H)$ value. In the literature there are two distinct formulations for these indicators (Pettini & Pagel 2004; Marino et al. 2013); in this work, we provide metallicity estimates using both of these indicators.

To compare the metallicity with the values found for the host galaxies of LGRB with an associated SN presented in Japelj et al. 2018, we used also the prescriptions developed by Maiolino et al. 2008 (M08 in Table 4). We consider the value of $12 + \log(O/H) = 8.69$ for the Solar metallicity (Asplund et al. 2009). All the values found are reported in Table 4. Finally, with the use of the [S II] $\lambda\lambda6717/6732$ doublet we can estimate the ionisation level, i.e. the ratio of the ionising photons vs. the particle density (Díaz & Pérez-Montero 2000). We obtain for the GRB region the value of $\log U_{GRB} = -2.60 \pm 0.18$ while for the entire host $\log U_{host} = -2.84 \pm 0.14$.

The metallicity found for the GRB region is similar to the typical values found by Japelj et al. 2018 for the regions of Type Ic-BL SNe with an associated GRB and for long GRB hosts in general (Vergani et al. 2017).

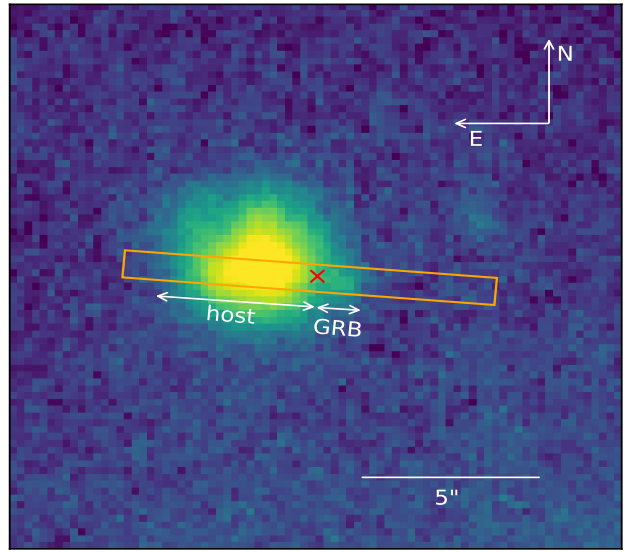

**Figure 7.** LBT $r$-band image of the host galaxy of GRB 171010A. The radio position of the GRB is marked with a red cross (Laskar et al. 2017). The orange box corresponds to the position of the X-shooter slit (length = 11″, width = 0.9″), whose configuration allows us to study the properties of both the GRB and the host galaxy.

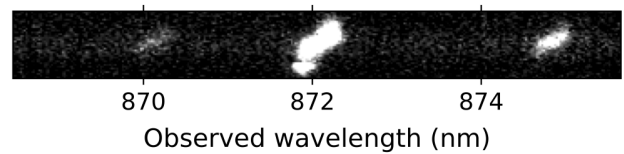

**Figure 8.** The 2D spectrum obtained with X-shooter showing the region around the H$\alpha$ $\lambda6562$ line, in the observer frame. The [N II] $\lambda\lambda6548,6584$ lines are also visible. The GRB position corresponds to the bright spot visible only in H$\alpha$, suggesting a low metallicity of the region at the GRB location.

**Table 4.** Host galaxy and GRB region emission lines measurements. The fluxes are in $10^{-17}$ erg cm$^{-2}$ s$^{-1}$ units.

| Emission line | GRB | host galaxy |
|---|---|---|
| [OII] 3727 | 2.02± 0.16 | 9.14 ± 0.18 |
| [OII] 3729 | 2.72 ± 0.14 | 18.70 ± 0.19 |
| H$\beta$ | 3.32 ± 0.16 | 13.00 ± 0.22 |
| [OIII] 4959 | 4.57 ± 0.15 | – |
| [OIII] 5007 | 15.30 ± 0.20 | 9.02 ± 0.23 |
| H$\alpha$ | 11.40 ± 0.24 | 53.60 ± 0.27 |
| [NII] 6584 | 0.77 ± 0.19 | 17.10 ± 0.24 |
| [SII] 6717 | 1.16 ± 0.24 | 11.9 ± 0.26 |
| [SII] 6732 | 0.80 ± 0.28 | 8.55 ± 0.31 |

**Table 5.** Physical properties of the GRB region and of the host galaxy covered by the X-shooter slit, inferred from emission line indicators. The second and third columns show the values of the SFR, as obtained from the Hα emission line (Kennicutt et al. 1994) and from the [O II] doublet (Kewley et al. 2004). The metallicity (12 + log(O/H)) measured using the indicators described in the text are reported in the fourth to eight columns, while in the last column the ionization level is listed (Díaz & Pérez-Montero 2000). The N2 and O3N2 tabulated values do not include the additional uncertainties inferred from the calibration of the relations (0.16 and 0.18 dex, respectively).

|  | SFR(Hα) ($M_\odot yr^{-1}$) | SFR([OII]) ($M_\odot yr^{-1}$) | N2 (M13) | O3N2 (M13) | N2 (PP04) | O3N2 (PP04) | M08 | log U |
|---|---|---|---|---|---|---|---|---|
| GRB | 0.19±0.04 | 0.24±0.07 | 8.20±0.05 | 8.14±0.05 | 8.23±0.06 | 8.14±0.03 | $8.30^{+0.06}_{-0.08}$ | -2.60±0.18 |
| Host (part of) | 0.88±0.04 | 1.42±0.43 | 8.51±0.02 | 8.46±0.05 | 8.62±0.01 | 8.62±0.01 | $8.94^{+0.05}_{-0.06}$ | -2.84±0.14 |
| GRB+host | 1.07±0.18 | 1.69±0.25 | 8.48±0.02 | 8.38±0.02 | 8.57±0.04 | 8.50±0.03 | $8.80^{+0.04}_{-0.05}$ | -2.82±0.15 |

## 4 CONCLUSIONS

We have presented a detailed study of SN 2017htp, a Type Ib/c core-collapse supernova associated with the long GRB 171010A at z = 0.33. We analysed the optical photometry and spectroscopy of GRB 171010A and SN 2017htp spanning nearly four months since its discovery. The supernova is confirmed both photometrically and spectroscopically and this event represents an example of the GRB-SN connection at moderately high redshift (Cano et al. 2017; Klose et al. 2019). We find that 0.33 $M_\odot$ of nickel is required to reproduce the peak luminosity of SN 2017htp, with an ejecta mass of $M_{ej}$ = 4.1 ± 0.7 $M_\odot$ and a kinetic energy of $E_K$ = 8.1 ± 2.5 × $10^{51}$ erg. Those values are consistent with previously observed GRBs-SNe.

We studied in detail the properties of the GRB region and the part of the host galaxy covered by the X-shooter spectroscopy. With a diameter of about half of that of the Milky Way (and a similar visible absolute magnitude), the host galaxy is among the biggest GRB hosts, second only to the GRB 060505 host galaxy, lacking a SN association and whose short or long nature is still debated (Thöne et al. 2008; Kann et al. 2011). We estimated the star-formation rate and metallicity of the GRB region with different line indicators. The observed properties of the GRB star-forming region (SFR$_{GRB}$ ∼ 0.2 $M_\odot$ yr$^{-1}$; 12 + log(O/H) ∼ 8.15 ± 0.10) are similar to those of the star-forming regions hosting other GRBs with an associated Type Ic-BL SN and with available spatially resolved observations (Christensen et al. 2008; Krühler et al. 2017; Izzo et al. 2017; Japelj et al. 2018). Long GRBs progenitor-star models and host galaxy observations indicate that long GRBs avoid high-metallicity environments (Vergani et al. 2015; Japelj et al. 2016; Perley et al. 2016; Palmerio et al. 2019). Nonetheless, a small fraction of long GRB host galaxies show high-metallicity. The observations of the host galaxy of GRB 171010A support the scenario that the metallicity of the GRB environment is low even in high-metallicity host galaxies.

## ACKNOWLEDGEMENTS

We thank the anonymous referee for the valuable comments that contributed to improving the quality of the publication. AM, PDA, SCa and GT acknowledge support from ASI grant *INAF I*/004/11/3. DBM is supported by research grant 19054 from Villum Fonden. Partly based on data acquired under the extended Public ESO Spectroscopic Survey for Transient Objects (ePESSTO; http://www.pessto.org) and within the STARGATE collaboration. Based on observations collected at the European Organisation for Astronomical Research in the Southern Hemisphere under ESO programme 199.D-0143. LI and DAK acknowledge the support from the Spanish research project AYA2014-58381-P. LI and DAK acknowledges support from Juan de la Cierva Incorporación fellowship IJCI-2016-30940 and IJCI-2015-26153, respectively. Based on observations collected at the Centro Astronómico Hispano Alemán (CAHA) at Calar Alto, operated jointly by the Max-Planck Institut für Astronomie and the Instituto de Astrofísica de Andalucía (CSIC). AR acknowledges support from Premiale LBT 2013. Based on data collected with Large Binocular Cameras at the LBT. The LBT is an international collaboration among institutions in the United States, Italy, and Germany. Based on observations made with the Nordic Optical Telescope under program 51-504, operated by the Nordic Optical Telescope Scientific Association at the Observatorio del Roque de los Muchachos, La Palma, Spain, of the Instituto de Astrofísica de Canarias. Partially based on observations collected at Copernico Telescope (Asiago, Mt. Ekar, Italy) of the INAF Osservatorio Astronomico di Padova. LT is partially supported by the PRIN-INAF 2017 "Towards the SKA and CTA era: discovery, localisation and physics of transient sources". MN is supported by a Royal Astronomical Society Research Fellowship. MG is supported by the Polish NCN MAESTRO grant 2014/14/A/ST9/00121. KM acknowledges support from H2020 ERC grant no.58638. The Cosmic Dawn Center is funded by the DNRF. Support for FOE is provided by the Ministry of Economy, Development, and Tourism's Millennium Science Initiative through grant IC120009, awarded to The Millennium Institute of Astrophysics. FOE acknowledges support from the FONDECYT grant no.1170953. DX acknowledges the supports by the One-Hundred-Talent Program of the Chinese Academy of Sciences (CAS), and by the Strategic Priority Research Program "Multi-wavelength Gravitational Wave Universe" of the CAS (No. XDB23000000).

## APPENDIX A: AUTHORS AFFILIATIONS

[1] INAF - Osservatorio Astronomico di Brera, Via E. Bianchi 46, I-23807, Merate (LC), Italy
[2] Dark Cosmology Centre, Niels Bohr Institute, University of Copenhagen, Lyngbyvej 2, DK-2100 Copenhagen Ø, Denmark
[3] The Cosmic Dawn Center (DAWN)
[4] Instituto de Astrofísica de Andalucía (IAA-CSIC), Glorieta de la Astronomía s/n, E-18008, Granada, Spain
[5] Anton Pannekoek Institute for Astronomy, University of Amsterdam, Science Park 904, 1098 XH Amsterdam, The Netherlands
[6] GEPI, Observatoire de Paris, PSL University, CNRS, Place Jules Janssen, 92195, Meudon, France
[7] Max-Planck Institut für Extraterrestrische Physik, Giessenbachstrsse 1, D-85748 Garching, Germany
[8] Nordic Optical Telescope, La Palma, Canary Islands 3537, Spain
[9] Department of Astronomy, The Oskar Klein Center, Stockholm University, AlbaNova, 10691 Stockholm, Sweden
[10] Mullard Space Science Laboratory, University College London, Holmbury St Mary, Dorking, Surrey RH5 6NT, United Kingdom
[11] INAF - Osservatorio Astronomico di Roma, Via di Frascati, 33, I-00040, Monteporzio Catone, Italy
[12] ASI - Science Data Centre, Via del Politecnico snc, I-00133, Roma, Italy
[13] Istanbul University Department of Astronomy and Space Sciences, 34119 Beyazıd, Istanbul, Turkey
[14] Centre for Astrophysics and Cosmology, Science Institute, University of Iceland, Dunhagi 5, 107 Reykjavík, Iceland
[15] INAF - Osservatorio di Astrofisica e Scienza dello Spazio di Bologna, Via Piero Gobetti 93/3, I-40129, Bologna, Italy
[16] Astrophysics Research Centre, School of Mathematics and Physics, Queens University Belfast, Belfast BT7 1NN, UK
[17] Clemson University, Department of Physics and Astronomy, Clemson, SC 29634-0978, USA
[18] Departamento de Ciencias Físicas, Universidad Andrés Bello, Fernández Concha 700, Las Condes, Santiago, Chile
[19] INAF - Istituto di Astrofisica Spaziale e Fisica Cosmica, via A. Corti 12, I-20133, Milano, Italy
[20] Institute for Astronomy, University of Edinburgh, Royal Observatory, Blackford Hill, EH9 3HJ, UK
[21] Birmingham Institute for Gravitational Wave Astronomy and School of Physics and Astronomy, University of Birmingham, B15 2TT, UK
[22] European Southern Observatory, Alonso de Córdova 3107, Casilla 19, Santiago, Chile
[23] Warsaw University Astronomical Observatory, Al. Ujazdowskie 4, 00-478, Warszawa, Poland
[24] INAF - Osservatorio Astronomico di Capodimonte, Salita Moiariello 16, I-80131, Napoli, Italy
[25] Astrophysics Research Centre, School of Mathematics and Physics, QueenâĂŹs University Belfast, Belfast BT7 1NN, UK
[26] Astronomical Institute Anton Pannekoek, University ofAmsterdam, PO Box 94249, 1090 GE Amsterdam, the Netherlands
[27] Instituto de Astronomía y Ciencias Planetarias, Universidad de Atacama, Copayapu 485, Copiapó, Chile
[28] Millennium Institute of Astrophysics (MAS), Nuncio Monseñor Sótero Sanz 100, Providencia, Santiago, Chile
[29] School of Physics & Astronomy, Cardiff University, Queens Buildings, The Parade, Cardiff, CF24 3AA, UK
[30] European Southern Observatory, Karl-Schwarzschild-Strasse 2, D-85748 Garching bei Munchen, Germany
[31] INAF - Osservatorio Astronomico di Padova, Vicolo dell'Osservatorio 5, I-35122, Padova, Italy
[32] Department of Astronomy, Stockholm University, SE-10691 Stockholm, Sweden
[33] Niels Bohr Institute, University of Copenhagen, Lyngbyvej 2, DK-2100 Copenhagen Ø, Denmark
[34] DTU Space, National Space Institute, Technical University of Denmark, Elektrovej 327, 2800 Kongens Lyngby, Denmark
[35] Department of Physics and Astronomy, University of Turku, Vesilinnantie 5, FI-20014 Turku, Finland
[36] Finnish Centre for Astronomy with ESO (FINCA), University of Turku, Vesilinnantie 5, FI-20014 Turku, Finland
[37] CAS Key Laboratory of Space Astronomy and Technology, National Astronomical Observatories, Chinese Academy of Sciences, Beijing 100101, China

This paper has been typeset from a TeX/LaTeX file prepared by the author.